\documentstyle[preprint,tighten,aps]{revtex}

\begin{document}
\author{Sheng Li\thanks{%
lisheng@itp.ac.cn}}
\address{Institute of Theoretical Physics, Academia Sinica, Beijing 10080,
P. R. China}
\author{Yishi Duan\thanks{%
ysduan@lzu.edu.cn}}
\address{Institute of Theoretical Physics,
Lanzhou University, Lanzhou 730000,
P. R. China}
\title{The Bifurcation of the Topological Structure in the Sunspot's Electric
Topological Current with Locally Gauge-invariant Maxwell-Chern-Simons Term}
\date{August 27, 1998}
\draft
\maketitle

\begin{abstract}
The topological structure of the electric topological current of the locally
gauge invariant Maxwell-Chern-Simons Model and its bifurcation is studied.
The electric topological charge is quantized in term of winding number. The
Hopf indices and Brouwer degree labeled the local topological structure of
the electric topological current. Using $\Phi $-mapping method and implicity
theory, the electric topological current is found generating or annihilating
at the limit points and splitting or merging at the bifurcate points. The
total electric charge holds invariant during the evolution.
\end{abstract}

\section{Introduction}

In order to search for the model of sunspots to generate mass to the gauge
field, Saniga, and Klacka (1992a-d etc.) had given the so-called Abelian
Higgs (AH) model of sunspots, which are regarded as topological AH magnetic
vortices. The central idea of the model is that the confinement of sunspot's
magnetic field into a finite-dimensional, sharply-bounded flux tube is due
to the so-called Higgs mechanism; a Higgs field spontaneously breaks an
original U(1)-symmetry of a corresponding field configuration giving thus
rise to a non-zero mass of photon. As another crucial point in such approach
it turned out to be the assumption of cylindrical symmetry of sunspots.
However there exist another interesting possibility of generating mass to
the gauge field that might be seen as alternative to the standard Higgs
mechanism. Namely, a photon can acquire a mass if the Lagrangian density
contains, in addition to the classical Maxwellian term, also the so-called
Chern-Simons (CS) one (see, e.g. Moradi, 1992). Although the resulting
Lagrangian density leads to gauge-invariant equations of motion it is not
itself a gauge-invariant quantity. Later Saniga (1993) brought the
Lagrangian density to a gauge-invariant form and showed that it admits
topological vortices. These vortices were then examined in some detail and
they are shown to carry both the quantized magnetic flux and electric
charge; but also endowed with a non-zero angular momentum they might thus
serve as a starting point for the description of rotating sunspots.

The aim of the present paper is to reveal the topological properties of the
electric topological current which is given in the locally gauge-invariant
Chern-Simons model of the sunspot. By regarding the complex scalar field $%
\zeta $, induced in the locally gauge-invariant MCS theory, as the complex
representation of a two-dimensional vector field $\vec{\zeta}=(\zeta
^1,\zeta ^2)$ with $\zeta ^1$ and $\zeta ^2$ being the real and imaginary
part of $\zeta $, the electric current is expressed as a topological
current. Using the $\phi $-mapping method and generalized function theory,
the structure of the electric topological current is proved taking the same
form as the classical current density in usual hydrodynamics, in which the
point-like particle with topological charge $%
{\displaystyle {\Theta \pi  \over e}}%
\beta _l\eta _l$, which is just the winding number of $\zeta $ at its
zeroes, are called topological electric charge. The locally topological
structure of the electric topological current is detailed, which is
quantized by the winding number of the integral surface and the field
function $\zeta $ and is labeled by the Hopf indices and Brouwer degrees of $%
\zeta $. Further, by imposing the implicit theory, we show that there exist
the crucial case of branch process in the electric topological current. It
will be seen that the electric topological current is unstable and splitting
(or merging) at the points where the corresponding Jacobian determinant $D(%
\frac \zeta u)$ of field function $\zeta $ vanishes. The branch processes at
the limit points as well as at the bifurcation point in the electric
topological current are studied systemically. The electric topological
current is found generating or annihilating at the limit points and
splitting or merging at the bifurcation points. The former relates to the
origin of the electric topological current. For the electric topological
current is identically conserved, the total electric charge holds invariant
during the process.

\section{Maxwell-Chern-Simons Electromagnetism and Electric Topological
Current Density}

Under the Maxwell-Chern-Simons configuration of field in the
(2+1)-dimensional Minkowski space-time, the field configuration represented
by the following Lagrangian density 
\begin{equation}
{\cal L}^{\odot }=-\frac 14F_{\rho \sigma }F^{\rho \sigma }+\frac \Theta 4%
\epsilon ^{\mu \nu \rho }F_{\mu \nu }A_\rho   \label{lagrangian}
\end{equation}
where $F_{\rho \sigma }\equiv \partial _\rho A_\sigma -\partial _\sigma
A_\rho $, $\Theta $ is a constant, and the space-time is endowed with the
metric tensor of the signature $(+--)$. In order to generalize (\ref
{lagrangian}) in a way to exhibit a gauge-invariant behavior, the Lagrangian
is modified in the way 
\[
{\cal L}_G^{\odot }=-\frac 14F_{\rho \sigma }F^{\rho \sigma }+\frac \Theta 4%
\epsilon ^{\mu \nu \rho }F_{\mu \nu }[A_\rho -\frac i{2e\zeta \zeta ^{*}}%
\frac \partial {\partial x^\mu }(\zeta ^{*}\frac{\partial \zeta }{\partial
x^\nu }-\zeta \frac{\partial \zeta ^{*}}{\partial x^\nu })]
\]
with $\zeta $ having the following gauge transformation rule 
\[
\zeta \rightarrow \tilde{\zeta}=e^{i\alpha }\zeta ,\quad \quad \quad \zeta
^{*}\rightarrow \tilde{\zeta}^{*}=e^{-i\alpha }\zeta ^{*};
\]
no other constraints are imposed on the scalar, complex-valued field $\zeta $
except the requirement of its single-valuedness. Using the modified
Lagrangian, one can get the electric topological current density $j^\rho $
of matter field, i.e. 
\begin{equation}
j^\rho =-\frac{i\Theta }{4e}\epsilon ^{\mu \nu \rho }\frac \partial {%
\partial x^\mu }(\frac{\zeta ^{*}}{\zeta \zeta ^{*}}\frac{\partial \zeta }{%
\partial x^\nu }-\frac \zeta {\zeta \zeta ^{*}}\frac{\partial \zeta ^{*}}{%
\partial x^\nu })
\end{equation}
This expression is of a topological nature since it is conserved
automatically 
\[
\frac{\partial j^\rho }{\partial x^\rho }=-\frac{i\Theta }{4e}\epsilon ^{\mu
\nu \rho }\frac{\partial ^2}{\partial x^\rho \partial x^\mu }(\frac{\zeta
^{*}}{\zeta \zeta ^{*}}\frac{\partial \zeta }{\partial x^\nu }-\frac \zeta {%
\zeta \zeta ^{*}}\frac{\partial \zeta ^{*}}{\partial x^\nu })\equiv 0,
\]
due to a complete antisymmetricity of the Levi-Civit\`{a} symbol. Hence the
spot's total electric charge $Q^{el}$ 
\begin{equation}
Q^{el}\equiv \int j^\rho dS_\rho =-\int \frac{i\Theta }{4e}\epsilon ^{\mu
\nu \rho }\frac \partial {\partial x^\mu }(\frac{\zeta ^{*}}{\zeta \zeta ^{*}%
}\frac{\partial \zeta }{\partial x^\nu }-\frac \zeta {\zeta \zeta ^{*}}\frac{%
\partial \zeta ^{*}}{\partial x^\nu })dS_\rho   \label{charge}
\end{equation}
is a purely topological quantity in exactly the same way as postulated by
Kovner and Rosenstein (1992) for electric charge in quantum electrodynamics.

Because the complex scalar field $\zeta $ can be denoted by 
\begin{equation}
\zeta =\zeta ^1+i\zeta ^2,  \label{pff}
\end{equation}
we regard $\zeta $ as the complex representation of the vector field 
\begin{equation}
\vec{\zeta}=(\zeta ^1,\zeta ^2).  \label{fff}
\end{equation}
Then we can define an unit vector 
\begin{equation}
n^a=\zeta ^a/\parallel \zeta \parallel ,\,\,\,\,\,\,\,\parallel \zeta
\parallel =\sqrt{\zeta ^a\zeta ^a},  \label{nff}
\end{equation}
satisfying 
\[
n^an^a=1,\,\,\,\,\,\,\,\,\,\,a=1,2.
\]
In fact $\vec{n}$ is identified as a section of the sphere bundle over the
space-time, i.e. $\vec{n}$ is a section of the $U(1)$ line bundle. It is
obvious that the zeroes of $\vec{\zeta}$ are just the singular points of $%
\vec{n}$. Now $j^\rho $ can be written as 
\begin{equation}
j^\rho =\frac \Theta {2e}\epsilon ^{\mu \nu \rho }\epsilon ^{ab}\partial
_\mu n^a\partial _\nu n^b  \label{u1connection}
\end{equation}
This formula of $j^\rho $ takes the same form as the topological current of
the torsion in Duan, Zhang and Feng's paper (1994). Hence we  call $j^\rho $
electric topological current.

Substituting (\ref{u1connection}) into (\ref{charge}), the spot's total
electric charge is 
\begin{equation}  \label{mapping}
Q^{el}=\int_S\frac \Theta {2e}\epsilon ^{\mu \nu \rho }\epsilon
^{ab}\partial _\mu n^a\partial _\nu n^bdS_\rho
\end{equation}

\section{Quantization and Local Topological Properties of the Sunspot's
Electric Topological Current}

To study the quantization and local topological properties of the electric
topological current, let us choose coordinates $y=(u^1,u^2,\tau )$ of the
space-time such that $u=(u^1,u^2)$ be the intrinsic coordinate on $S$. For
the coordinate component $\tau $ does not belong to $S$. Then 
\begin{equation}
Q^{el}=\int_S\frac \Theta {2e}\epsilon ^{ij}\epsilon ^{ab}\frac{\partial n^a%
}{\partial u^i}\frac{\partial n^b}{\partial u^j}dS
\end{equation}
where $i,j=1,2$ and 
\begin{equation}
dS=\frac 12\epsilon ^{\upsilon v\rho }\epsilon _{ij}\frac{\partial u^i}{%
\partial x^\mu }\frac{\partial u^j}{\partial x^\nu }dS_\rho =du^1du^2
\end{equation}
is the element of the surface $S$. Notice 
\begin{equation}
\epsilon ^{ij}\epsilon ^{ab}\frac{\partial n^a}{\partial u^i}\frac{\partial
n^b}{\partial u^j}=\epsilon ^{ab}\epsilon ^{ij}\frac \partial {\partial
\zeta ^c}\frac \partial {\partial \zeta ^a}\ln \Vert \zeta \Vert \cdot \frac{%
\partial \zeta ^c}{\partial u^i}\frac{\partial \zeta ^b}{\partial u^j}%
dS_\rho .
\end{equation}
Define the Jacobian determinant $D(\frac \zeta u)$ as 
\begin{equation}
\epsilon _{ij}D(\frac \zeta u)=\epsilon ^{ab}\frac{\partial \zeta ^a}{%
\partial u^i}\frac{\partial \zeta ^b}{\partial u^j}.
\end{equation}
Using of Laplacian relation in $\zeta $-space 
\begin{equation}
\frac{\partial ^2}{\partial \zeta ^a\partial \zeta ^a}\ln \Vert \zeta \Vert
=2\pi \delta ^2(\zeta ),
\end{equation}
we rewrite the equation (\ref{mapping}) in a compact form 
\begin{equation}  \label{5}
Q^{el}=\frac{\Theta \pi }e\int D(\frac \zeta k)\delta ^2(\zeta )dS
\end{equation}
The equation (\ref{5}) shows that only those points, on which $\zeta =0$,
contribute to $Q^{el}$.

Suppose that the vector field $\zeta ^a(x)$ possesses $N$ zeroes on $S$ and
let the $l$th zero be $x=z_l$%
\begin{equation}
\zeta ^a(z_l)=0
\end{equation}
According to the deduction of Duan and Liu (1988) and the implicit function
theorem (see e.g. Edourd, 1904), the solutions of $\zeta (u^1,u^2,\tau )=0$
can be expressed in terms of $u=(u^1,u^2)$ as 
\begin{equation}
u^i=z^i(\tau ),\quad \quad \quad \quad \quad i=1,2
\end{equation}
and 
\begin{equation}
\zeta ^a(z_l^1(\tau ),z_l^2(\tau ),\tau )\equiv 0,  \label{zeropoint}
\end{equation}
where the subscript $l=1,2,\cdots ,N$ represents the $l$th zero of $\zeta $,
i.e. 
\begin{equation}
\zeta ^a(z_l^i)=0,
\end{equation}
It is easy to get the following formula from the ordinary theory of $\delta $%
-function that 
\begin{equation}
\delta ^2(\zeta )D(\frac \zeta u)=\sum_{l=1}^N\beta _l\eta _l\delta (u-z_l).
\end{equation}
The positive integer $\beta _l$ is called the Hopf index of map $%
x\rightarrow \zeta $ (see, e.g. Milnor, 1965; Dubrovin, 1985; Duan 1979),
which means that when the point $x$ covers the neighborhood of the zero $%
x=z_l$ once, the function $\zeta ^i$ covers the corresponding region in $%
\zeta $-space $\beta _l$ times. And 
\begin{equation}
\eta _l=\frac{D(\frac \zeta u)}{|D(\frac \zeta u)|}|_{x=z_l}=\pm 1,
\end{equation}
is called the Brouwer degree of map $x\rightarrow \zeta $ (see Duan 1979,
1990). That the Hopf indices be integers is due to the single-valueness of $%
\zeta $. Substituting this expansion of $\delta ^2(\zeta )$ into (\ref{5}), $%
Q^{el}$ is quantized in the topological level as 
\begin{equation}
Q^{el}=\frac{\Theta \pi }e\int \sum_{l=1}^N\beta _l\eta _l\delta ^2(u-z_l)dS=%
\frac{\Theta \pi }e\sum_{l=1}^N\beta _l\eta _l.  \label{7}
\end{equation}
Hence, the total electric charge of the MCS sunspot is quantized, i.e. it is
composed of an integer number of the `universal' unit $Q_0^{el}$.
Furthermore, from (\ref{7}) we see that this total electric charge is
composed by many independent quantized charges which locate at the zeroes of
the complex scalar field $\zeta $ and the topological structure of the
electric density is labeled by Hopf index $\beta _i$ and Brouwer degree $%
\eta _i$.

On another hand, the winding number of the surface $S$ and the mapping $%
\zeta $ is defined as (Victor and Alan, 1974) 
\begin{equation}
W=\frac 1{2\pi }\int_S\epsilon ^{ij}\epsilon ^{ab}\frac{\partial n^a}{%
\partial u^i}\frac{\partial n^b}{\partial u^j}dS.
\end{equation}
which is equal to the number of times $S$ encloses (or, wraps around) the
point $\zeta =0$. Hence, the total electric charge is quantized by the
winding number 
\begin{equation}
Q^{el}=W\frac{\Theta \pi }e.  \label{w-q}
\end{equation}
The winding number $W$ of the surface $S$ can be interpreted or, indeed,
defined as the degree of the mapping $\zeta $ onto $S$. By (\ref{5}) we have 
\begin{eqnarray*}
Q^{el} &=&\frac{\Theta \pi }e\int_S\delta (\zeta )D(\frac \zeta u)du^1du^2 \\
&=&\frac{\Theta \pi }e\deg \zeta \int_{\zeta (S)}\delta (\zeta )d\zeta
^1d\zeta ^2 \\
&=&\frac{\Theta \pi }e\deg \zeta
\end{eqnarray*}
where $\deg \zeta $ is the degree of map $\zeta :S\rightarrow \zeta (S)$.
Compared above equation with (\ref{w-q}), it shows the degree of map $\zeta
:S\rightarrow \zeta (S)$ is just the winding number $W$ of the surface $S$
and the mapping $\zeta $, i.e. 
\[
W=\deg \zeta 
\]
Then the total electric topological charge is 
\begin{equation}
Q^{el}=WQ_0^{el}=\deg \zeta Q_0^{el}.
\end{equation}
Divide $S$ by 
\[
S=\sum_{l=1}^NS_l 
\]
and $S_l$ includes only one zero $z_l$ of $\zeta $, i.e. $z_l\in S_l$. Then
The winding number of the surface $S_l$ and the mapping $\zeta $ is 
\[
W_l=\frac 1{2\pi }\int_{S_l}\epsilon ^{ij}\epsilon ^{ab}\frac{\partial n^a}{%
\partial u^i}\frac{\partial n^b}{\partial u^j}dS 
\]
which is equal to the number of times $S_l$ encloses (or, wraps around) the
point $u=z_l$. It is easy to see that 
\[
W=\sum_{l=1}^NW_l 
\]
and 
\[
\beta _l=|W_l|\quad \quad \quad \quad \eta _l=signW_l. 
\]
Then 
\[
Q^{el}=Q_0^{el}\sum_{l=1}^NW_l=Q_0^{el}\sum_{l=1}^N\beta _l\eta _l. 
\]

Denote the sum of $\beta _l$ with $\eta _l=1$ and $\eta _l=-1$ as $W^{+}$
and $W^{-}$ respectively, the total electric charge can be rewritten as 
\begin{equation}
Q^{el}=\frac{\Theta \pi }e(W^{+}-W^{-}),  \label{last-hcon}
\end{equation}
which reveal the contributions of electric topological current with positive
or negative charges. Since the zero of $\zeta $ at $u=z_l$ gives a
vortexlike structure, the expression (\ref{charge}) gives the vorticity in
space-time. So $W^{+}$ and $W^{-}$ denote the vorticity of vortex and
antivortex. Therefore the vortex and antivortex classify positive or
negative charges.

\section{The Bifurcation of the Topological Structure of Electric
Topological Current}

Using the intrinsic coordinates, redefine the topological electronic current
as 
\begin{equation}
j^\rho =\frac \Theta {2e}\epsilon ^{\mu \nu \rho }\epsilon ^{ab}\frac{%
\partial n^a}{\partial y^\mu }\frac{\partial n^b}{\partial y^v}
\end{equation}
From (\ref{zeropoint}) we can prove that the general velocity of the $l$th
zero 
\begin{equation}  \label{velocity}
V^j:=\frac{dz_l^j}{d\tau }=\frac{D^j(\zeta /u)}{D(\zeta /u)}|_{u=z_l}\quad
\quad \quad V^0=1.
\end{equation}
Then the electric topological current $j^\rho $ can be written as the form
of the current density of the system of $l$ classical point particles with
topological charge $Q_l^{el}=\eta _l\beta _lQ_0^{el}$ moving in the ($2+1$%
)-dimensional space-time 
\begin{equation}  \label{f564}
j^i=\sum\limits_{l=1}^NQ_l^{el}\delta (u-z_l(\tau ))\frac{du_l^i}{d\tau }%
,\qquad j^0=\sum\limits_{l=1}^NQ_l^{el}\delta (k-z_l(\tau ))
\end{equation}
The total charge of the system is 
\begin{equation}  \label{f565}
Q^{el}=\int_Sj^\rho dS_\rho =\int_Sj^0dS=Q_0^{el}\sum\limits_{l=1}^N\beta
_l\eta _l,
\end{equation}
Therefore we get a concise expression for the topological current, 
\begin{equation}  \label{f567}
j^i=j^0\frac{D^i(\zeta /u)}{D(\zeta /u)}=j^0V^i=Q_0^{el}\sum_{l=1}^N\beta
_l\eta _lV^i,
\end{equation}
which takes the same form as the current density in hydrodynamics.
Expressions (\ref{f564}) and (\ref{f567}) give the topological structure of
the electric topological current, which is characterized by the Brouwer
degrees and Hopf indices. In our theory the point-like particles with
topological charges $Q^{el}_l=Q_0^{el}\beta _l\eta _l\,(l=1,2,\cdots ,N)$
are called topological particles, the charges of which are topologically
quantized and these particles are just located at the zeros of $\zeta (u)$,
i.e. the singularities of the unit vector $n(u)$.

The above discussion is based on the condition that the Jacobian 
\begin{equation}  \label{nonzero}
D(\frac \zeta u)|_{z_l}\neq 0.
\end{equation}
When $D(\frac \zeta u)|_{z_l}=0$, it is shown that there exist the crucial
case of branch process. There are two kinds of branch points namely limit
points and bifurcation points, which will be discussed in detail in the
follows.

Firstly we study the case when the zeros of $\vec{\zeta}$ include some limit
points. The limit points are determined by 
\begin{equation}
\left\{ 
\begin{array}{l}
\zeta ^1(u^1,u^2,\tau )=0 \\ 
\zeta ^2(u^1,u^2,\tau )=0 \\ 
\zeta ^3(u^1,u^2,\tau )=D(\frac \zeta u)=0
\end{array}
\right.  \label{88}
\end{equation}
and 
\begin{equation}
D^A(\frac \zeta u)|_{(z_l,\tau ^{*})}\neq 0,\quad \quad A=1,2  \label{89}
\end{equation}
where we denote the limit points as $(z_l,\tau ^{*})$. Since the usual
implicit function theorem is of no use when the Jacobian determinant $D(%
\frac \zeta u)|_{z_l}=0$. For the purpose of using the implicit function
theorem to study the branch properties of electric topological current at
the limit points, we use the Jacobian $D^1(\frac \zeta y)$ instead of $D(%
\frac \zeta u)$ to search for the solutions of $\zeta =0$. This means we
have replaced $u^1$ by $\tau $. For clarity we rewrite the first two
equations of (\ref{88}) as 
\begin{equation}
\zeta ^a(\tau ,u^2,u^1)=0,\quad \quad a=1,2.  \label{91}
\end{equation}
Taking account of (\ref{89}) and using the implicit function theorem, we
have a unique solution of the equations (\ref{91}) in the neighborhood of
the limit point $(z_l,\tau ^{*})$%
\begin{equation}
\tau =\tau (u^1),\quad \quad u^2=u^2(u^1)  \label{92}
\end{equation}
with $\tau ^{*}=\tau (z_l^1)$. In order to show the behavior of the electric
topological current at the limit points, we will investigate the Taylor
expansion of (\ref{92}) in the neighborhood of $(z_l,\tau ^{*})$. In the
present case, from (\ref{89}) and the last equation of (\ref{88}), we get 
\[
\frac{du^1}{d\tau }|_{(z_l,\tau ^{*})}=\frac{D^1(\frac \zeta u)}{D(\frac %
\zeta u)}|_{(z_l,\tau ^{*})}=\infty 
\]
i.e. 
\[
\frac{d\tau }{du^1}|_{(z_l,\tau ^{*})}=0. 
\]
Then, the Taylor expansion of $\tau =\tau (u^1)$ at the limit point $%
(z_l,\tau ^{*})$ is 
\[
\tau =\tau (u_l^1)+\frac{d\tau }{du^1}|_{(z_l,\tau ^{*})}(u^1-z_l^1)+\frac 12%
\frac{d^2\tau }{(du^1)^2}|_{(z_l,\tau ^{*})}(u^1-z_l^1)^2 
\]
\begin{equation}
=\tau ^{*}+\frac 12\frac{d^2\tau }{(d\tau ^1)^2}|_{(z_l,\tau
^{*})}(u^1-z_l^1)^2.
\end{equation}
Therefore 
\begin{equation}
\tau -\tau ^{*}=\frac 12\frac{d^2\tau }{(du^1)^2}|_{(z_l,\tau
^{*})}(u^1-z_l^1)^2  \label{93}
\end{equation}
which is a parabola in $u^1-\tau $ plane. From (\ref{93}) we can obtain two
solutions $u_1^1(\tau )$ and $u_2^1{}(\tau )$, which give the branch
solutions of electric topological current at the limit points. If $\frac{%
d^2\tau }{(du^1)^2}|_{(z_l,\tau ^{*})}>0$, we have the branch solutions for $%
\tau >\tau ^{*}$, otherwise, we have the branch solutions for $\tau <\tau
^{*}$. The former related to the origin of the topological electric charges
and the later is related the annihilate of the topological electric charges.

Since the electric topological current is identically conserved, the
topological quantum numbers of these two generated or annihilated
topological electric charges must be opposite at the limit point, i.e. 
\[
\beta _1\eta _1Q_0^{el}+\beta _2\eta _2Q_0^{el}=0, 
\]
which shows that the limit points do not contribute to the total electric
charge.

Now, let us turn to consider the other case, in which the restrictions are 
\begin{equation}  \label{det-bif}
D(\frac \zeta u)|_{(z_l,\tau ^{*})}=0,\quad \quad \quad D^1(\frac \zeta u%
)|_{(z_l,\tau ^{*})}=0.
\end{equation}
These two restrictive conditions will lead to an important fact that the
function relationship between $\tau $ and $u^1$ is not unique in the
neighborhood of $(\tau ^{*},z_l)$. In our electric topological current
theory this fact is easily seen from one of the Eqs.(\ref{velocity}) 
\begin{equation}  \label{v-bif}
V^1=\frac{du^1}{d\tau }=\frac{D^1(\frac \zeta u)}{D(\frac \zeta u)}%
|_{(z_l,\tau ^{*})}
\end{equation}
which under (\ref{det-bif}) directly shows that the direction of the
integral curve of (\ref{v-bif}) is indefinite at $(z_l,\tau ^{*})$.
Therefore the very point $(z_l,\tau ^{*})$ is called a bifurcation point of
the electric topological current. With the aim of finding the different
directions of all branch curves at the bifurcation point, we suppose that 
\begin{equation}  \label{phi-k2}
\frac{\partial \zeta ^1}{\partial u^2}|_{(z_l,\tau ^{*})}\neq 0.
\end{equation}
From $\zeta ^1(u^1,u^2,\tau )=0$, the implicit function theorem says that
there exists one and only one function relationship 
\begin{equation}  \label{k2-k1}
u^2=u^2(u^1,\tau ).
\end{equation}
Substituting (\ref{k2-k1}) into $\zeta ^1$, we have 
\[
\zeta ^1(u^1,u^2(u^1,\tau ),\tau )\equiv 0 
\]
which gives 
\begin{equation}  \label{phi-f}
\frac{\partial \zeta ^1}{\partial u^2}f_1^2=-\frac{\partial \zeta ^1}{%
\partial u^1},\quad \quad \quad \frac{\partial \zeta ^1}{\partial u^2}f_\tau
^2=-\frac{\partial \zeta ^1}{\partial \tau },
\end{equation}
\[
\frac{\partial \zeta ^1}{\partial k^2}f_{11}^2=-2\frac{\partial ^2\zeta ^1}{%
\partial k^2\partial k^1}f_1^2-\frac{\partial ^2\zeta ^1}{(\partial u^2)^2}%
(f_1^2)^2-\frac{\partial ^2\zeta ^1}{(\partial u^1)^2}, 
\]
\[
\frac{\partial \zeta ^1}{\partial u^2}f_{1\tau }^2=-\frac{\partial ^2\zeta ^1%
}{\partial u^2\partial \tau }f_1^2-\frac{\partial ^2\zeta ^1}{\partial
u^2\partial u^1}f_\tau ^2-\frac{\partial ^2\zeta ^1}{(\partial k^2)^2}f_\tau
^2f_1^2-\frac{\partial ^2\zeta ^1}{\partial k^1\partial \tau }, 
\]
\[
\frac{\partial \zeta ^1}{\partial u^2}f_{\tau \tau }^2=-2\frac{\partial
^2\zeta ^1}{\partial u^2\partial \tau }f_\tau ^2-\frac{\partial ^2\zeta ^1}{%
(\partial u^2)^2}(f_\tau ^2)^2-\frac{\partial ^2\zeta ^1}{\partial \tau ^2}, 
\]
where the partial derivatives is 
\[
f_1^2=\frac{\partial u^2}{\partial u^1},\quad \quad f_\tau ^2=\frac{\partial
u^2}{\partial \tau },\quad \quad f_{11}^2=\frac{\partial ^2u^2}{(\partial
u^1)^2},\quad \quad 
\]
\[
f_{1\tau }^2=\frac{\partial ^2u^2}{\partial u^1\partial \tau },\quad \quad
f_{\tau \tau }^2=\frac{\partial ^2u^2}{\partial \tau ^2}. 
\]
From these expressions it is easy to calculate the values of $f_1^2,f_\tau
^2,f_{11}^2,f_{1\tau }^2$ and $f_{\tau \tau }^2$ at $(\tau ^{*},z_l)$.

In order to explore the behavior of the electric topological current at the
bifurcation points, let us investigate the Taylor expansion of 
\begin{equation}
f(u^1,\tau )=\zeta ^2(u^1,u^2(u^1,\tau ),\tau )  \label{f-phi2}
\end{equation}
in the neighborhood of $(z_l,\tau ^{*})$, which according to the Eqs.(\ref
{88}) must vanish at the bifurcation point, i.e. 
\begin{equation}
f(z_l,\tau ^{*})=0.  \label{4.51}
\end{equation}
From (\ref{f-phi2}), the first order partial derivatives of $f(u^1,\tau )$
with respect to $u^1$ and $\tau $ can be expressed by 
\begin{equation}
\frac{\partial F}{\partial u^1}=\frac{\partial \zeta ^2}{\partial u^1}+\frac{%
\partial \zeta ^2}{\partial u^2}f_1^2,\quad \quad \quad \frac{\partial F}{%
\partial \tau }=\frac{\partial \zeta ^2}{\partial \tau }+\frac{\partial
\zeta ^2}{\partial u^2}f_\tau ^2.  \label{fk-ft}
\end{equation}
By making use of (\ref{phi-f}), (\ref{fk-ft}) and Cramer's rule, it is easy
to prove that the two restrictive conditions (\ref{det-bif}) can be
rewritten as 
\[
D(\frac \zeta u)|_{(z_l,\tau ^{*})}=(\frac{\partial f}{\partial u^1}\frac{%
\partial \zeta ^1}{\partial u^2})|_{(z_l,\tau ^{*})}=0, 
\]
\[
D^1(\frac \zeta u)|_{(z_l,\tau ^{*})}=(\frac{\partial f}{\partial \tau }%
\frac{\partial \zeta ^1}{\partial u^2})|_{(z_l,\tau ^{*})}=0, 
\]
which give 
\begin{equation}
\frac{\partial f}{\partial u^1}|_{(z_l,\tau ^{*})}=0,\quad \quad \quad \frac{%
\partial f}{\partial \tau }|_{(z_l,\tau ^{*})}=0  \label{4.53}
\end{equation}
by considering (\ref{phi-k2}). The second order partial derivatives of the
function $f$ are easily to find out to be 
\begin{eqnarray*}
\frac{\partial ^2f}{(\partial u^1)^2} &=&\frac{\partial ^2\zeta ^2}{%
(\partial u^1)^2}+2\frac{\partial ^2\zeta ^2}{\partial u^2\partial u^1}f_1^2+%
\frac{\partial \zeta ^2}{\partial u^2}f_{11}^2+\frac{\partial ^2\zeta ^2}{%
(\partial u^2)^2}(f_1^2)^2 \\
\frac{\partial ^2f}{\partial u^1\partial \tau } &=&\frac{\partial ^2\zeta ^2%
}{\partial u^1\partial \tau }+\frac{\partial ^2\zeta ^2}{\partial
u^2\partial u^1}f_\tau ^2+\frac{\partial ^2\zeta ^2}{\partial u^2\partial
\tau }f_1^2+\frac{\partial \zeta ^2}{\partial u^2}f_{1\tau }^2+\frac{%
\partial ^2\zeta ^2}{(\partial u^2)^2}f_1^2f_\tau ^2 \\
\frac{\partial ^2f}{\partial \tau ^2} &=&\frac{\partial ^2\zeta ^2}{\partial
\tau ^2}+2\frac{\partial ^2\zeta ^2}{\partial u^2\partial \tau }f_\tau ^2+%
\frac{\partial \zeta ^2}{\partial u^2}f_{\tau \tau }^2+\frac{\partial
^2\zeta ^2}{(\partial u^2)^2}(f_\tau ^2)^2
\end{eqnarray*}
which at $(z_l,\tau ^{*})$ are denoted by 
\begin{equation}
A=\frac{\partial ^2f}{(\partial u^1)^2}|_{(z_l,\tau ^{*})},\quad B=\frac{%
\partial ^2f}{\partial u^1\partial \tau }|_{(z_l,t^{*})},\quad C=\frac{%
\partial ^2f}{\partial \tau ^2}|_{(z_l,\tau ^{*})}.  \label{4.54}
\end{equation}
Then, from (\ref{4.51}), (\ref{4.53}) and (\ref{4.54}), we obtain the Taylor
expansion of $f(u^1,\tau )$ 
\[
f(u^1,\tau )=\frac 12A(u^1-a_l^1)^2+B(u^1-a_l^1)(\tau -\tau ^{*})+\frac 12%
C(\tau -\tau ^{*})^2 
\]
which by (\ref{f-phi2}) is the behavior of $\zeta ^2$ in the neighborhood of 
$(z_l,t^{*})$. Because of the second equation of (\ref{88}), we get 
\[
A(u^1-a_l^1)^2+2B(u^1-a_l^1)(\tau -\tau ^{*})+C(\tau -\tau ^{*})^2=0 
\]
which leads to 
\begin{equation}
A(\frac{du^1}{d\tau })^2+2B\frac{du^1}{d\tau }+C=0  \label{4.55}
\end{equation}
and 
\begin{equation}
C(\frac{d\tau }{du^1})^2+2B\frac{d\tau }{du^1}+A=0.  \label{4.56}
\end{equation}
The solutions of equations (\ref{4.55}) or (\ref{4.56}) give the different
directions of the branch curves at the bifurcation point. The remainder
component $du^2/d\tau $ can be given by 
\[
\frac{du^2}{d\tau }=f_1^2\frac{du^1}{d\tau }+f_\tau ^2 
\]
where partial derivative coefficients $f_1^2$ and $f_\tau ^2$ have been
calculated in (\ref{phi-f}). Now we get all the different directions of the
branch curves. This means the behavior of the electric topological current
at the bifurcation points is detailed.

The above solutions reveal the evolution of the electric topological
current. Besides the generation and the annihilation of the topological
charges, it also includes the spliting and merging of electric topological
current. When an original electric topological current moves through the
bifurcation point in $k$-space, it may split into two electric topological
currents moving along different branch curves, or,two original electric
topological currents merge into one electric topological current at the
bifurcation point. The identical conversation of the electric topological
current shows the total electric charge must keep invariant before and after
the evolution at the bifurcation point, i.e. 
\[
\beta _{l_1}\eta _{l_1}+\beta _{l_2}\eta _{l_2}=\beta _l\eta _l 
\]
for fixed $l$. Furthermore, from above studies, we see that the generation,
annihilation and bifurcation of electric topological current are not gradual
changes, but start at a critical value of arguments, i.e. a sudden change.

\section{Summary and Conclusion}

In this paper, with the gauge potential decomposition and the so called $%
\zeta $-mapping method, we obtain the inner topological structure of
electric topological current and its evolution. The topological quantization
of electric topological current is gotten. The electric topological current
is found to take a form of generalized function $\delta (\zeta )$ $%
\zeta=(\zeta ^1,\zeta ^2 )$ with $\zeta ^1,\zeta ^2$ are the real part and
imaginary part of the magnetic wave function. The Hopf indices $\beta _l$
and Brouwer degree $\eta _l$ of the magnetic wave function reveal the inner
topological structure of electric topological current, which are also bring
to light the effect of vortices with different current-carrying property.
When the Jacobian $D(\frac \zeta u)=0$, it is shown that there exist the
crucial case of branch process. Based on the implicit function theorem and
the Taylor expansion, the evolution of the electric topological current is
detailed in the neighborhoods of the branch points of $\zeta $-mapping. The
branch solutions at the limit points and the different directions of all
branch curves at the bifurcation points are calculated out. Because the
electric topological current is identically conserved, the total electric
charge will keep to be a constant during the evolution. At the limit points
case, it means that the topological quantum numbers of the two generated or
annihilated electric topological currents must be opposite at the limit
point. It can be looked upon as the topological origin of electric
topological charge. At the bifurcation case, the total charges of different
branch electric topological currents keeps invariant, which is the
topological reason of the conservation law.

\acknowledgements
This research was supported by National Natural Science Foundation of P. R.
China.

\section*{References}

\noindent {}\hskip -5mmDuan, Y.S. and Ge, M.L.: 1979,{\em \ Sci. Sinica., }%
{\bf 11}, 1072.

\noindent \hskip -5mmDuan, Y.S., and Zhang, S.L., and Feng, S.S.: 1994, {\em %
J. Math. Phys.}{\it \ }{\bf 35, }1.

\noindent \hskip -5mmDuan, Y.S. and Liu, J.C.: 1988, in {\em Proceedings of
Johns Hopkins Workshop 11}, edited by Duan, Y.S., Domokos G. and
Kovesi-Domokos, S., World Scientific, Singapore.

\noindent \hskip -5mmDuan, Y.S. and Zhang, S.L.: 1990, {\em Int. J. Eng. Sci.%
}\ {\bf 28} 689.

\noindent \hskip -5mmDubrovin, B.A., et.al. (1985). {\em Modern Geometry
Method and Application}, Part II, Springer-Verlag New York Inc.


\noindent \hskip -5mm\'{E}douard Goursat,: 1904, {\em A Course in
Mathematical Analysis, Vol.I,} translated by Earle mond Hedrick.

\noindent \hskip -5mmKovner, A and Rosenstein, B.: 1992, {\em Int. J. Mod.
Phys. }{\bf A7}, 7419.

\noindent \hskip -5mmMarandi, G: 1992, {\em The role of Topology in
Classical and quantumn Physics, Springer-Verlag}, Heidelberg.

\noindent \hskip -5mmMilnor J. W.: 1965, {\em Topology, From the
Differential Viewpoint} The University Press of inebreak Virginia
Charlottesville.

\noindent \hskip -5mmSaniga, M.: 1992a, {\em Astron. Zh.} {\bf 69},902.

\noindent \hskip -5mmSaniga, M.: 1992b, {\em Astron. Space Sci.} {\bf 194},
229.

\noindent \hskip -5mmSaniga, M.: 1992c, {\em Astron. Space Sci.} {\bf 197},
109.

\noindent \hskip -5mmSaniga, M.: 1992d, {\em Astron. Space Sci.} {\bf 193},
155.

\noindent \hskip -5mmSaniga, M.: 1993c, {\em Astron. Space Sci.} {\bf 207},
119.

\noindent \hskip -5mmVictor G. and Alan P.: 1974, {\em Differential Topology}%
, Prentice-Hall, Inc., Englewood Cliffs, New Jersey.

\end{document}